\documentclass[10pt,twocolumn,a4paper]{article}

\usepackage[T1]{fontenc}
\usepackage[utf8]{inputenc}
\usepackage{lmodern}

\usepackage{amsmath,amssymb,amsfonts}

\usepackage{graphicx}
\usepackage{booktabs}
\usepackage{multirow}
\usepackage{xurl}
\usepackage{hyperref}

\usepackage[font=small]{caption}

\usepackage{lipsum}

\usepackage[margin=1.8cm]{geometry}

\title{Energy Storage as a Multi-Use Asset: \\Applications Across the Power System}
\author{Fabrizio Sossan\\
HES-SO Valais-Wallis\\
\texttt{fabrizio.sossan@hevs.ch}}
\date{\today}

\begin{document}

\maketitle

\begin{abstract}
The energy transition in power systems requires flexible assets to offset renewable generation variability across multiple time scales, while supporting the integration of renewables and the electrification of demand without requiring costly grid reinforcement. Energy storage occupies a unique position among these assets: depending on the technology, it can provide short-duration grid services at high ramping rates, such as frequency regulation and voltage support, longer-duration functions such as intra-day peak shaving, or inter-seasonal energy buffering. This multi-service character, combined with the declining costs of energy storage technologies (most notably that of battery energy storage systems), is central to the economic viability of storage investments. The value of a given installation depends strongly on its grid connection point and intended use case: an asset-coupled battery serving a consumer or generation plant faces a different service landscape, and therefore a different business case, than a network-coupled system operating as an independent grid resource. This paper presents a structured taxonomy of grid-connected energy storage applications, discusses the principal application domains, and describes the key challenges that must be addressed to integrate storage effectively into power systems. Services are discussed with special emphasis on the Swiss regulatory context. Finally, the STORE flagship project supported by the Swiss Innovation Agency (Innosuisse), where some of the critical challenges of energy storage integration in power grids are addressed, is introduced.
\end{abstract}

\begingroup
\renewcommand{\thefootnote}{}
\footnotetext{The support of the Swiss Innovation Agency Innosuisse through its Flagship programme under grant agreement 108.230~FS-EE (STORE) is gratefully acknowledged.}
\footnotetext{This manuscript has been submitted to the Grid Services \& Markets Conference (GSM 2026).}
\addtocounter{footnote}{-1}
\endgroup

\section{Introduction}

The energy transition underway in many countries is reshaping power systems fundamentally. According to the Swiss Federal Office of Energy, the Swiss power grid may host up to 30 GW of installed PV capacity by 2040 \cite{SFOE2022}. To put this figure in perspective, this amounts to almost double the current installed hydropower capacity (approximately 19 GW), a historically critical energy infrastructure in Switzerland that today accounts for more than 60\% of yearly Swiss electricity consumption.
This transition poses challenges at multiple levels. At the distribution grid level, infrastructure originally designed for unidirectional power flow may not be capable of hosting large amounts of distributed PV generation, a problem amplified by the simultaneous electrification of transport and heating. At the transmission and market operation level, the system operator, as load-balance responsible, faces more variable and less predictable generation patterns, with shorter forecast horizons and larger imbalances. At the generation level, existing dispatchable plants may not be capable of providing the ramping rates and cycling frequencies that a high-renewable system requires, or may incur significantly higher maintenance costs in attempting to do so. Finally, the progressive replacement of conventional synchronous generation with power-converter-interfaced resources reduces system inertia, requiring a fundamental revision of existing stability assessment frameworks and new strategies to compensate for the reduced short-circuit current capacity of converter-based generation. Taken together, these challenges point toward a common need: flexible assets capable of absorbing variability, restoring controllability, and supporting grid infrastructure across multiple time scales. Energy storage is a technology that can address some of these needs effectively.
Energy storage is often considered an enabler for integrating larger shares of renewable generation, thanks to its controllable and dispatchable power output that can be leveraged for multiple services, from mitigating renewable variability to supporting grid operations. The current interest in energy storage is motivated by the confluence of several factors: the growth of renewables (especially PV generation in Switzerland) is challenging the business-as-usual operations of both grid operators and generation companies; the rapidly decreasing costs of battery energy storage systems (BESS), driven by declining battery pack prices \cite{BloombergNEF2025}, have made these technologies accessible to a broad range of operators; and the resulting investment activity is further supported by attractive electricity market conditions, such as capacity payments in ancillary service markets.
In this context, many operators are looking with interest at BESS either to improve their operational performance or to access new revenue streams, and often both simultaneously. A key aspect in improving the payback time of energy storage systems is the capability of providing multiple services simultaneously, which can improve revenue stacks, shorten return on investment periods, and attract investments.
This paper presents a taxonomy of the services that energy storage can provide in power grids, both in current regulatory settings with special emphasis on the Swiss one, and in future perspectives. It then discusses the methodological challenges faced by operators planning and operating such systems, and describes selected application examples in detail. The paper concludes with a presentation of the STORE flagship project, supported by the Swiss Innovation Agency (Innosuisse), in which four academic and thirteen industrial partners jointly address pressing challenges related to the large-scale integration of renewable energy in power systems.

\section{A Taxonomy of Energy Storage Services and Applications}
\begin{figure*}[!ht]
    \centering
    \includegraphics[width=1\linewidth]{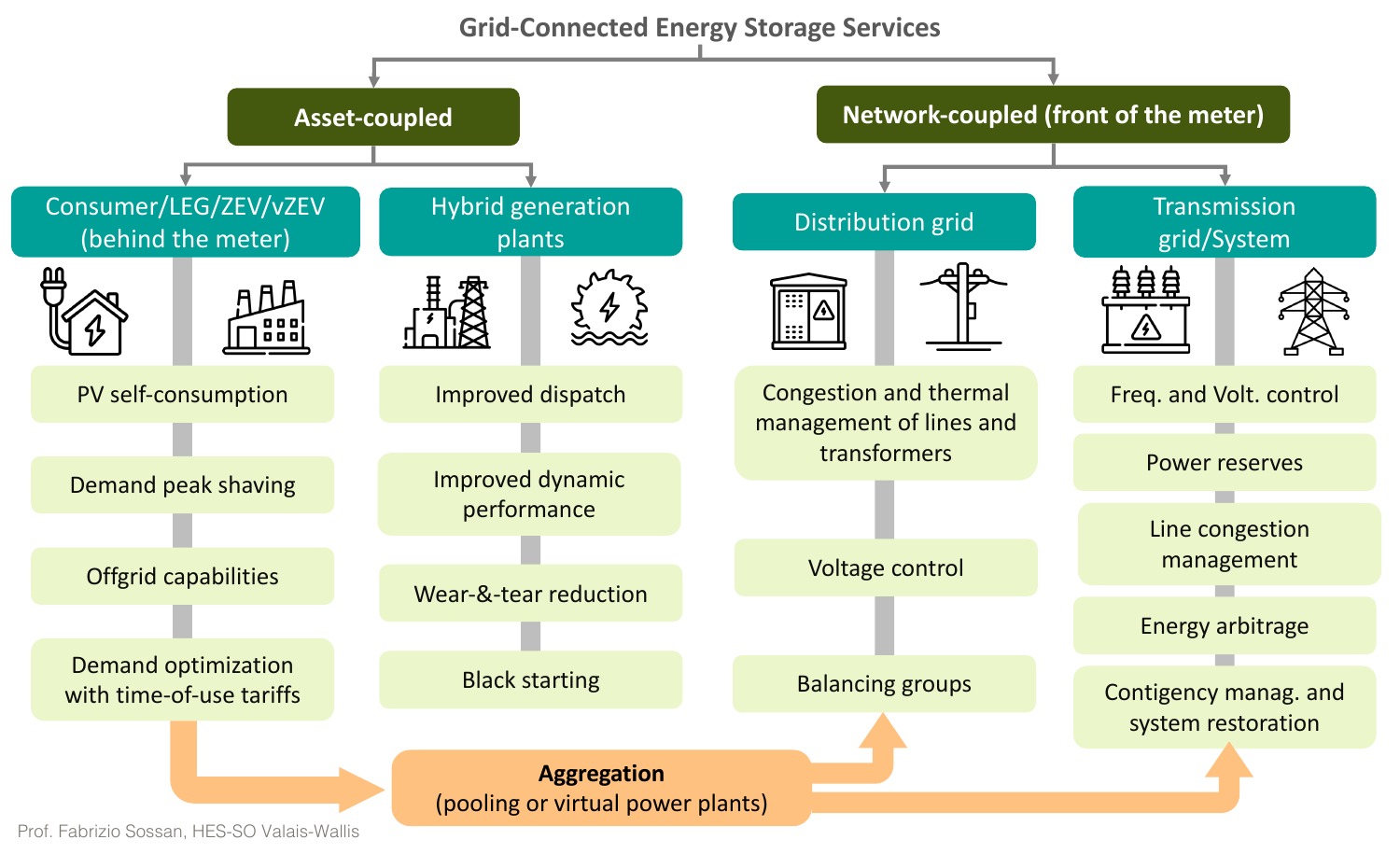}
    \caption{A taxonomy of Energy Storage Services and Applications.}
    \label{fig:taxonomy}
\end{figure*}

Fig.~\ref{fig:taxonomy} illustrates a taxonomy of energy storage services by grid level and primary function. 

The primary split divides between systems that are deployed alongside a consumer installation or generation plant and serve that asset's operational and economic objectives as its primary purpose, and assets that are installed at a generic point of the network standalone. Services for each category are described next.

\subsection{Consumer level (behind the meter)}
At the consumer level, the key services that energy storage systems can provide include:

\begin{itemize}
    \item \textbf{PV self-consumption maximization}. This application is primarily motivated by the reduction in feed-in tariffs for PV generation, which are typically lower than retail electricity tariffs for consumption. Swiss energy regulations foresee that, if no agreement is reached between a distribution system operator (DSO) and a PV producer, injected electricity is remunerated according to a quarterly reference market price calculated as the average day-ahead exchange price in the Swiss market zone, weighted by the actual quarter-hourly injection profile. Since January 2026, this reference price has also served as the nationally harmonized fallback remuneration mechanism when DSOs and producers cannot agree on injection conditions. Energy storage systems can therefore increase the economic value of PV generation by shifting local production toward local consumption.
    \item \textbf{Demand peak shaving}. Energy storage systems can reduce peak power consumption and thereby decrease the power-related component of the electricity bill. In addition, peak shaving may help avoid reinforcement of the grid connection point, the cost of which may, under Swiss energy legislation, be charged by the DSO to the final consumer if the upgrade primarily benefits that specific user.
    \item \textbf{Off-grid capability and backup power}. Energy storage systems can provide backup power in the event of grid outages, thereby increasing reliability of supply. In the case of critical processes or sensitive loads (e.g., in the chemical industry, data centers, or critical infrastructure), blackouts may incur very high economic costs. BESS, particularly when coupled with grid-forming converters, can provide resilience against outages and support islanded operation. Virtually all modern residential and commercial BESS are, for example, capable of autonomous backup operation. However, for larger sites requiring multiple controllable assets, or hybrid systems combining different controllable resources such as local hydropower units, the coordination of parallel grid-forming units requires careful design, including transient stability analysis and proper static load-sharing behavior through droop control, as well as integration with higher-level energy management and load-shedding strategies.
    \item \textbf{Demand optimization under time-of-use (ToU) tariffs}. Dynamic electricity prices may be used by utilities as an indirect mechanism to shift electricity consumption toward off-peak periods. Energy storage systems, together with schedulable flexible demand, can be controlled automatically in response to dynamic pricing signals, thereby implicitly supporting grid operation while reducing consumer electricity costs.
\end{itemize}
	
Behind-the-meter applications are today naturally extended toward energy communities and aggregation mechanisms, which are discussed later in \ref{sec:beyond_btm}.

\subsection{Hybrid generation plants}
This application refers to coupling generation plants with energy storage systems. The key services enabled by such hybridization are:

\begin{itemize}
    \item \textbf{Improved dynamic performance and ecological constraints}. Traditional power plants do not always respond instantaneously to control signals due to the dynamics of internal control loops as well as mechanical and thermal time constants. As a result, they may struggle to comply with the timing requirements imposed by modern grid codes. Energy storage systems, most notably BESS, are capable of adapting their power output very rapidly due to the absence of significant mechanical and thermal inertia, and can therefore help improve the real-time response of these units. Improved dynamic performance may also encompass aspects that are not directly related to the electrical domain. One example is the mitigation of hydropeaking in hydropower plants, where BESS can be used as a partial replacement for fast turbine activation, thereby smoothing discharge variations and helping preserve river ecology.

    \item \textbf{Improved dispatchability}. A local energy buffer can help improve dispatch trajectories, for example by reducing start-stop sequences in hydropower plants or by assisting in meeting spinning reserve requirements. For stochastic generation sources, a local energy buffer can support generation firming, reduce imbalance penalties in electricity markets, and provide a degree of controllability beyond classical curtailment strategies.

    \item \textbf{Wear-and-tear reduction}. Power plants, particularly hydropower plants in Switzerland, may experience significantly increased mechanical fatigue and wear due to frequent actuation of regulation organs such as guide vanes. This results in higher mechanical stress levels, increased cycling, and greater operational mileage (e.g., \cite{Cassano2022,Biner2022,Kadam2023}). Energy storage systems can alleviate these effects by taking over fatigue-intensive regulation duties. In this context, state-of-charge management and energy management can be performed seamlessly at the hybrid power plant level by redispatching local assets.

    \item \textbf{Black start capability}. Local energy storage, if operated while maintaining a residual energy margin, can provide the energy required to support black-start operations following outages. Due to the limited short-circuit current capability of converter-based systems, special care should be devoted to electromagnetic transient analysis of local auxiliary systems that may require high magnetizing currents, such as power transformers or induction motors (e.g., pumps).
\end{itemize}

It is worth highlighting that some of the benefits described above can also be achieved without physically co-locating the energy storage asset with the generation plant. For example, a company operating geographically dispersed assets may find it more advantageous to install storage at a central location and optimize operations at the fleet level. Nevertheless, co-location of storage with generation assets may be desirable in order to reuse existing infrastructure and avoid additional grid connection investments. This is particularly relevant when existing transformers with spare capacity, grid interconnection equipment, or DC bus infrastructure (variable speed with BESS) can be reused.

\subsection{Distribution grid}

The use of energy storage in distribution grids is typically intended as a non-wire reinforcement measure. Traditional reinforcement measures in distribution grids consist of replacing cables with higher-capacity ones, reconductoring existing lines, and upgrading transformers. Non-wire reinforcement generally refers to the use of alternative measures, commonly grouped under the term ``flexibility'', such as energy storage, flexible demand, and PV curtailment. The Swiss energy legislation requires DSOs to consider optimization measures, including the use of flexibility resources (such as storage), before proceeding with traditional grid reinforcement or grid expansion. Regulatory aspects are discussed in more detail in \label{sec:law} of this paper.

A properly operated energy storage system connected to the distribution grid can contribute to the following services:
\begin{itemize}
    \item \textbf{Congestion and thermal management}. Congestion management refers to ensuring that currents in lines and transformers do not exceed the ampacity limits of grid components during periods of overload. These limits typically reflect steady-state thermal constraints. In addition, temperature variations and thermal cycling may accelerate the aging of insulation materials and other grid assets, potentially causing premature component failures, and thus management of the thermal limits (thermal management).
    \item \textbf{Voltage control}. Voltage control refers to the requirement for operators to maintain voltage levels within normative limits (e.g., nominal value $\pm$10\% as per EN 50160). Voltage levels outside these limits may damage grid-connected equipment or cause tripping and malfunction of connected apparatus (e.g., brownouts). Voltage violations are also a common source of involuntary disconnection of PV converters. 
    \item \textbf{Balancing groups}. Energy storage can be used as a resource to compensate for deviations from balancing group schedules, thereby reducing balancing costs for grid operators.
\end{itemize}

Several approaches can be envisaged for using energy storage systems as non-wire alternatives in distribution grids for grid support services:

\begin{itemize}
    \item \textit{Incentivized behind-the-meter installations}. The DSO incentivizes the installation of behind-the-meter storage systems while retaining partial controllability over the assets. This model is currently being explored in the STORE project together with the grid operator RELL.
    \item \textit{Dynamic electricity pricing}. Energy storage assets may be steered through dynamic electricity prices computed to relieve grid congestion. Energy storage and other flexible resources are particularly well suited to this approach, as they can be controlled automatically using optimization-based control methods such as economic model predictive control (eMPC).
    \item \textit{Dedicated flexibility markets}. Dedicated flexibility markets for distribution grid operations could provide incentives for distributed resources to support grid operation. Such markets, however, do not currently exist in Switzerland.
    \item \textit{DSO-owned energy storage systems}. The DSO directly owns and operates the BESS. This option, however, is relatively uncommon and is generally only permissible if the storage asset is not used for activities outside the DSO domain (e.g., no energy arbitrage) due to the need of respecting separation of concerns among grid operators.
\end{itemize}

\subsection{Transmission grid}

Services that energy storage assets can provide to transmission system operators (TSOs) include:

\begin{itemize}
    \item \textbf{Frequency and voltage control}. Energy storage systems can contribute to frequency and voltage regulation across multiple time scales, ranging from very fast dynamics (synthetic inertia and primary frequency control) to slower services such as secondary and tertiary reserves. For fast time scales, BESS are generally excellent candidates due to their high ramping capability and fast response times. Slower services, particularly tertiary control, are typically more energy-intensive because of longer activation durations and may therefore be less suitable for short-duration storage technologies. Power converters can emulate inertial response and are generally four-quadrant devices, meaning they can also provide reactive power support for voltage regulation.

    \item \textbf{Power reserves}. Energy storage systems can act as standby resources capable of injecting or absorbing power when needed. Due to the remuneration mechanisms associated with reserve markets, which often compensate both reserved capacity and actual activation, these services currently provide strong revenue streams for BESS operators. This has significantly accelerated investments in the sector and fostered the deployment of large-scale energy storage installations.

    \item \textbf{Transmission congestion management}. Energy storage systems can help relieve congestion in high-voltage and extra-high-voltage transmission lines by locally absorbing or injecting power in order to maintain line currents within ampacity limits.

    \item \textbf{Energy arbitrage in electricity markets}. Energy arbitrage is a classical application of energy storage that leverages price differentials across electricity and ancillary service markets. By charging during low-price periods and discharging during high-price periods, storage systems can both generate profit and contribute to reducing peak demand, thereby limiting the activation of expensive peak-generation units. In systems with high penetration of renewable energy sources, storage can also help mitigate market volatility caused by the inherent intermittency of renewable generation and its impact on the merit order of electricity markets.

    \item \textbf{Contingency support and system restoration}. Energy storage assets with sufficient state-of-charge reserves can be rapidly rescheduled to cope with contingencies, thereby buying time for more coordinated and larger-scale corrective actions. In addition, storage systems can contribute to system restoration processes by providing energization points within the grid, which is a critical step during black-start and grid re-energization procedures.
\end{itemize}

One important challenge for front-of-the-meter energy storage systems is the supply chain, both for BESS and high-voltage grid connection equipment, as well as the authorization and permitting processes, which, in Switzerland, may require several years \cite{Swissgrid2025Batteries}. In this respect, smaller systems installed within distribution grids may present a competitive advantage, as they can often reuse existing grid infrastructure and avoid the need for additional transmission-level connection assets. On the other hand, the engineering and operational complexity associated with coordinating and controlling a large number of distributed systems in a reliable manner is generally higher than that of operating a single centralized BESS.

\subsection{Beyond behind-the-meter: Energy communities and clustering}\label{sec:beyond_btm}
An extension of behind-the-meter energy storage systems is the coordinated operation of generation, demand, and storage across multiple consumers and prosumers. In Switzerland, this evolution is reflected in new regulatory frameworks that define \cite{EnV2025,Mantelerlass2023}:

\begin{itemize}
    \item Collective self-consumption associations (ZEV in German, RCP in French and Italian)
    \item Virtual collective self-consumption associations (vZEV, vRCP)
    \item Local electricity communities (LEG, CEL)
\end{itemize}

\begin{figure*}[!ht]
    \centering
    \includegraphics[width=0.85\linewidth]{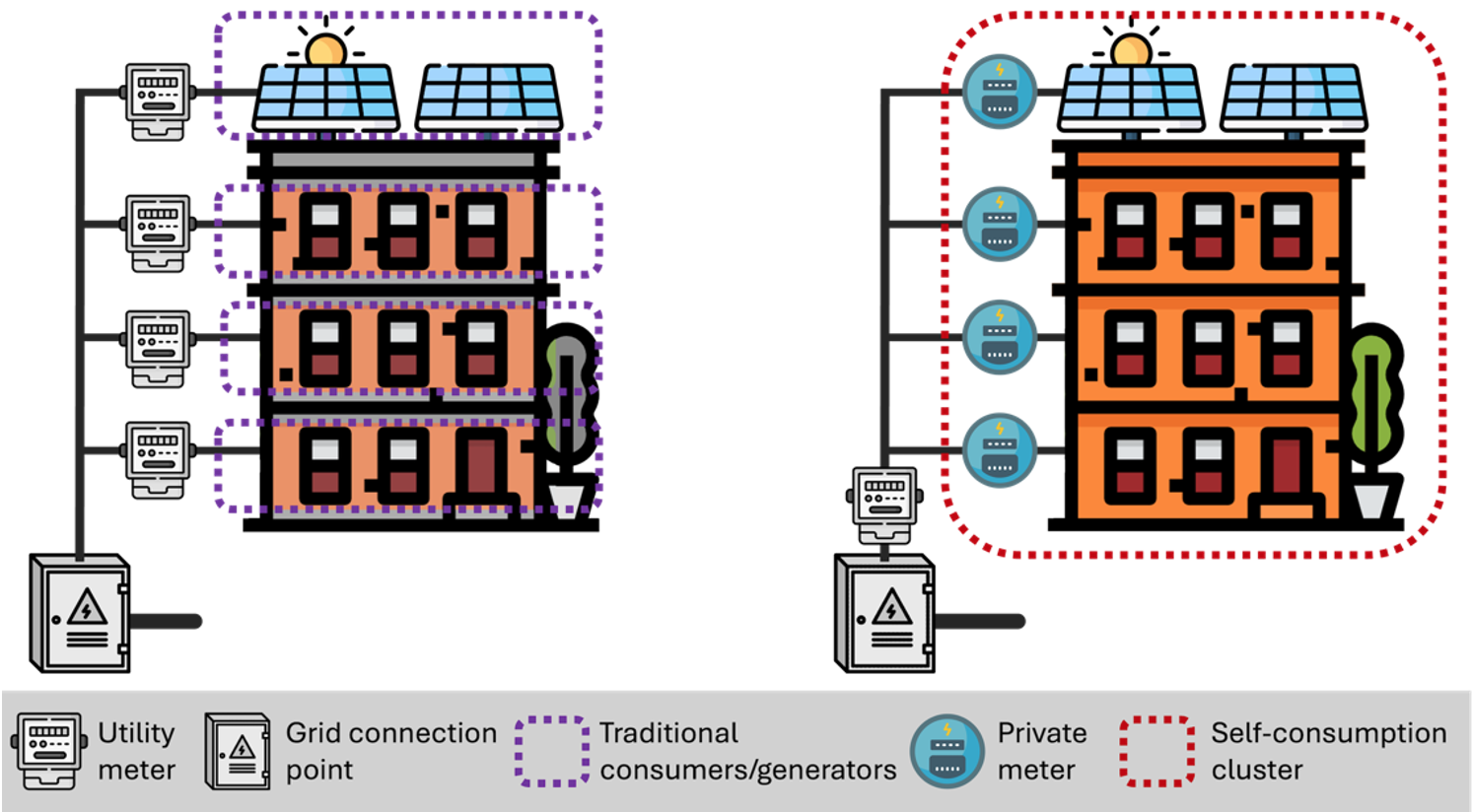}
    \caption{Left: individual billing points behind a common point of delivery. Right: individual billing points are aggregated into a collective self-consumption association (ZEV or RCP). Further extensions include virtual ZEVs (vZEVs), which span multiple points of delivery connected to the same LV feeder.}
    \label{fig:rcp}
\end{figure*}

\begin{figure*}[!ht]
    \centering
    \includegraphics[width=0.85\linewidth]{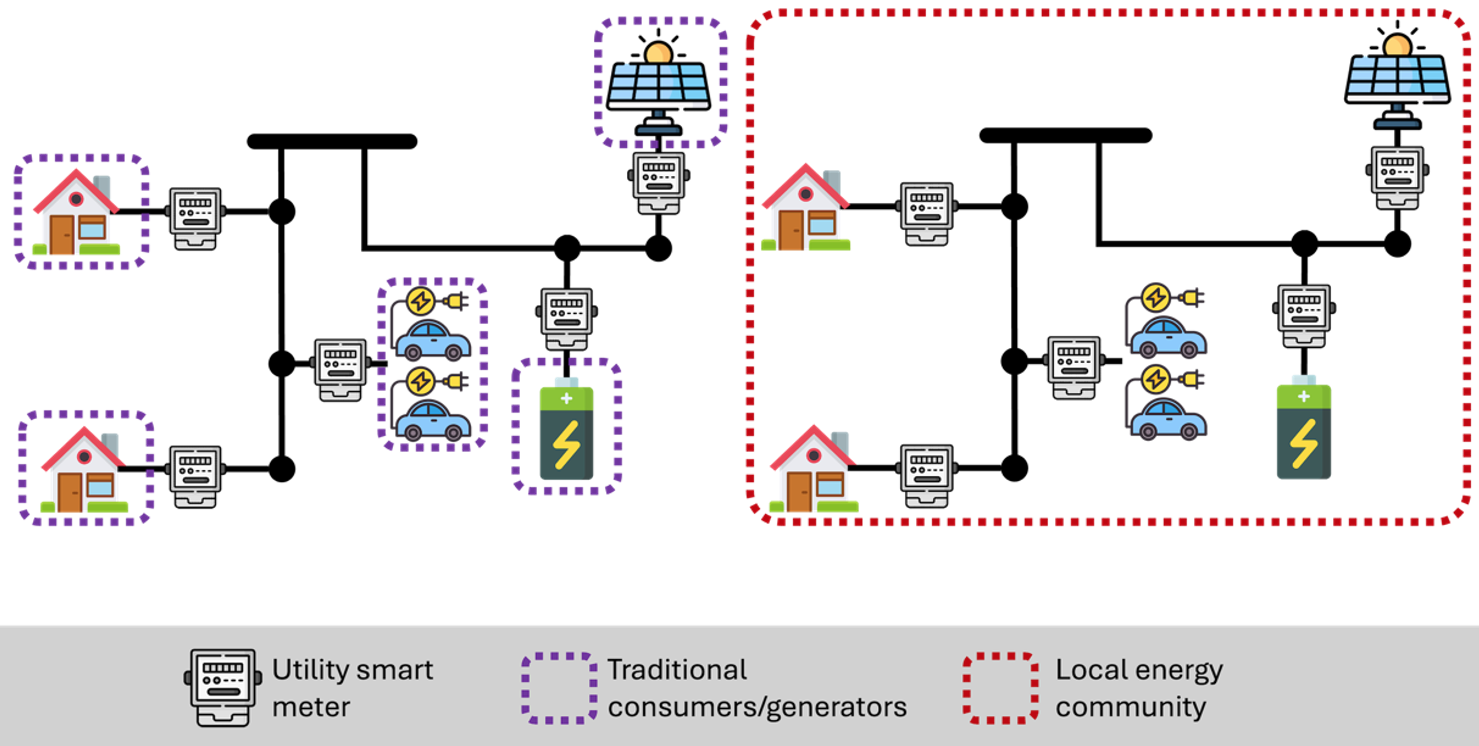}
    \caption{Left: individual points of delivery in a distribution grid; Right: aggregation into an energy community under an unified billing mechanism to favor self-consumption (LEG or CEL).}
    \label{fig:cel}
\end{figure*}

These frameworks, shown in figures \ref{fig:rcp} and \ref{fig:cel}, enable users to share locally generated renewable electricity and flexibility resources. A ZEV allows multiple consumers and producers located behind a common point of delivery (e.g., within the same building or site) to share locally produced electricity, typically PV generation, under a unified metering and billing structure. By avoiding multiple individual metering points and associated grid tariffs, significant cost savings may be achieved, especially when combined with energy storage systems. 
The vZEV framework, introduced in Switzerland in 2025, extends this concept across multiple points of delivery connected to the same low-voltage feeder. Local electricity communities (LEG/CEL) further generalize this approach by enabling broader aggregation structures that may span larger portions of the distribution network, including multiple feeders and voltage levels, while still incentivizing geographically and electrically local energy exchanges through reduced network tariffs (up to 60\%).

These aggregation mechanisms are particularly relevant for energy storage systems. Community-level storage (if properly managed with specialized energy management system) can increase PV self-consumption, reduce demand peaks, support local congestion and voltage management, and enable coordinated participation in flexibility services. In this context, storage operates not only as an individual consumer asset, but also as a shared flexibility resource whose value depends on the collective operation of the community.

Beyond fostering local PV self-consumption, which is the primary objective of such constructs, these frameworks may also constitute a natural aggregation layer, more specifically, an aggregation based on electrical network topology, that could support the provision of ancillary services to system operators. At the same time, they may foster greater social awareness of energy-related challenges and strengthen community engagement in the energy transition.

\subsection{Aggregation and virtual power plants}
A transversal dimension connecting small-scale assets (for example in behind-the-meter applications) with transmission-grid-level services (most notably ancillary services such as secondary and tertiary frequency control) is aggregation. Individual storage assets are often too small to participate directly in wholesale electricity or ancillary service markets, whose minimum bid sizes are typically on the order of several megawatts and therefore far exceed the capacity of typical behind-the-meter or local energy community (LEG) assets.
Aggregation addresses this limitation by pooling the capacity and flexibility of multiple distributed assets into a single market-facing entity. This practice is often referred to as a virtual power plant (VPP). A flexibility aggregator or VPP operator schedules the combined charge and discharge of its portfolio to meet market obligations, dispatching individual assets in real time through automated control signals. From the market’s perspective, the aggregated portfolio behaves as a single large resource; from the asset owners’ perspective, participation in the VPP generates revenue streams from services that would otherwise be inaccessible.
Software platforms and energy management systems capable of aggregating the resources of an energy community with high reliability and guaranteed performance, while simultaneously providing both behind-the-meter and front-of-the-meter services, currently constitute an active field of research and early industrial development. These systems offer significant potential by enabling access to both behind-the-meter and front-of-the-meter service markets.
Aggregation of energy communities, and aggregation within energy communities, may also provide quantitative and operationally meaningful solutions to the problem of exporting flexibility from distribution networks toward transmission system operators while respecting distribution-grid operational constraints. In this respect, topology-aware aggregation and DSO-coordinated flexibility management constitute complementary pathways toward the broader challenge of “DSO-safe” aggregation of distributed resources connected to low- and medium-voltage distribution networks \cite{SwissgridTSODSO2026}.

\subsection{Regulatory aspects related to flexibility and reinforcement in distribution grids in Switzerland}\label{sec:law}
Distribution grids and DSOs are at the cornerstone of the energy transition, as most innovative technologies, both new loads such as electric vehicles and heat pumps, and distributed generation such as rooftop PV, are typically connected at low- and medium-voltage levels due to their rated power and utilisation profile. In this context, Swiss legislation has evolved significantly in recent years to promote the utilisation of new technologies \cite{StromVG2026, StromVV2026}.
Swiss energy legislation requires DSOs to consider optimisation measures, including the use of flexibility, before proceeding with network reinforcement or expansion, while leaving the final assessment to the DSO (StromVG, Art.~9b). Flexibility use can be remunerated through a contract between the DSO and the flexibility provider (StromVV, Art.~19b), and should be managed through appropriate monitoring and control systems. The DSO is eligible to receive compensation for testing innovative “smart-grid” solutions, including the use of flexibility (StromVV, Art.~13b).
As far as grid reinforcement is concerned, costs associated with the connection of renewable generation are partially offset through a national contribution (CHF 59/kW) reimbursed by the national grid operator (Swissgrid) to the DSO \cite{StromVV2026}, with any residual connection costs charged to the producer (StromVV, Art.~15b). Upgrade costs associated with large PV installations without self-consumption may be directly attributed to the requesting customer (StromVV, Art.~16). DSOs are permitted to curtail PV generation when this improves operational conditions (StromVG, Art.~17c); however, curtailment exceeding 3\% of the annual yield must be financially compensated to the producer (Änderung der Stromversorgungsverordnung mit
Inkrafttreten am 1. Januar 2026).

\section{Integrating energy storage: from design to real-time control}
Integrating energy storage systems into grid operations is still a relatively new practice for grid and asset operators; standardized procedures for their design and operation are not yet fully established. The process typically starts from a specific need, or multiple ones, that justifies the deployment of a BESS. These needs are generally associated with a technical issue, an economic objective, or, in many cases, both. The project must then be developed to ensure that the designed system is capable of meeting the required specifications and operational objectives for which it was designed.
From a methodological perspective, putting into operation an energy storage application can be studied through three sequential and closely interconnected layers: design (or planning), scheduling, and application-level real-time control. The example here below refers to a BESS.

\subsection{Design (or planning)}
This phase aims at selecting the technology and determining the size of the system in terms of rated power (MVA) and energy capacity (MWh) required to deliver the targeted application. Technology selection should consider asset costs, degradation characteristics, including both calendar aging and cycling degradation and as a function of the services to be delivered. In stationary applications, energy and power density are generally much less critical than in electric vehicle applications.
Energy storage applications can be broadly classified as energy-intensive or power-intensive depending on whether the primary requirement is the amount of energy to be delivered or the rate at which it must be delivered. Examples of energy-intensive applications include PV self-consumption, which typically requires storing energy over several hours, with seasonal energy storage being even more energy intensive. Examples of power-intensive applications include primary frequency control or fatigue reduction in hydropower plants.
For power-intensive applications, a key parameter of an energy storage technology is its maximum achievable C-rate. Technologies with low C-rates require relatively large energy capacities to provide high power, forcing operators to install significant amounts of energy capacity even when power capability is the primary objective. Conversely, technologies with high C-rates can provide large power outputs with comparatively smaller energy capacities.
Nowadays, technology selection in stationary applications is largely driven by the market dominance of lithium iron phosphate, which offers a favorable trade-off among performance, safety, lifetime, and, especially, cost. However, future markets may offer alternative technologies, such as sodium-ion batteries or next-generation flow batteries, potentially providing advantages over current mainstream technologies in terms of C-rate capability, degradation characteristics, sustainability, or material availability.
In grid applications, the location of the storage system is also a critical parameter, as some nodes may have greater influence than others, for example with respect to voltage regulation or congestion management. This is known as the siting problem. Typical non-technical constraints arising in the siting problem include land-use constraints and safety considerations, given that many energy storage technologies may pose hazards such as fire risk or environmental contamination in the event of accidents. Access to a grid connection and the available hosting capacity are, of course, also critical aspects.

\subsection{Scheduling (energy management)}
Once a BESS is installed and operational, the first operational layer to address is energy management and scheduling, whose objective is to ensure that sufficient energy remains available in real time to deliver the intended services. Energy management problems are typically formulated as optimization problems, with state-of-the-art approaches relying on deterministic, stochastic, robust, or distributionally robust formulations to account for uncertainty. Because battery operation is intrinsically intertemporal through the state-of-charge dynamics, scheduling decisions must explicitly account for future operating conditions over finite prediction horizons. This generally requires the use of forecasts, such as numerical weather predictions, forecasts of future dispatch plans, or electricity and ancillary service market signals. Scheduling problems are typically solved again periodically with updated information so as to account for unfolding uncertainty and correct the SOC trajectory of the BESS.
Forecasts may take the form of point predictions, probabilistic forecasts, or scenario-based representations. Since many commercial BESS installations provide relatively limited energy buffers, often in the range of one to four hours, properly accounting for these intertemporal couplings is essential to ensure that the system can deliver the required performance when needed. Failing to do so, for example through myopic control strategies without predictive capabilities or through poorly calibrated forecasts, may significantly degrade operational performance and compromise the intended objectives of the system.

\subsection{Application-level real-time control}
The application-level real-time control layer is responsible for collecting real-time measurements and governing the BESS to meet the requirements of the designed application and the dispatch setpoints provided by the scheduler. By implementing and tracking the scheduler's instructions, it ensures that real-time performance requirements are met while the scheduled plan is followed, avoiding driving the BESS to full depletion or full charge.
This layer can be understood as a soft real-time control; it sends active and, if required by the application, reactive power setpoints to the BESS via a communication interface, commonly Modbus TCP, EtherCAT, or CANbus. Setpoints are typically refreshed at rates ranging from a few hundred milliseconds to tens of seconds depending on the application. These setpoints are processed by the power converter, which, by way of its internal control loops operating at substantially faster time scales, actuates the reference power to be delivered to the grid, provided that sufficient energy is available in the battery stack. Anything requiring faster control action or deterministic computation deadlines, such as primary frequency response, islanding detection, or droop control, must be handled directly by the converter or programmed into it at commissioning time, since the latency and non-deterministic timing of application-level communication interfaces make them unsuitable for such functions.
The application-level real-time control is not concerned with implementing internal BESS functionalities, such as cell balancing, thermal management, state-of-charge estimation, and protection against overcurrent and overvoltage, as these are managed autonomously by the BMS. This separation of concerns is deliberate: the BMS provides a safety backstop that remains active regardless of the state of the higher-level controller, ensuring safe operation even in the event of a failure or misbehavior at the application level. Active communication with the BMS is nonetheless necessary to read the operational status of the BESS, including the state of charge and the instantaneous power limits, which can change dynamically as a function of state of charge and temperature, and to propagate any errors or alarms to the upper-level monitoring system.
Finally, the operations of both the application-level real-time control and the scheduling layer should be monitored continuously, including through dedicated performance reports, for performance assessment and auditing purposes. Connection to a SCADA system is a non-negotiable prerequisite for coordinated operation: it allows the asset to receive dispatch instructions, report its state in real time, and remain visible to the energy management systems producing valid schedules. When the BESS operates alongside other assets, such as a hydropower plant, control and measurement signals must be integrated within the same supervisory system to ensure coherent operation. If applicable, the control system must also be capable of receiving external setpoints from aggregators and dispatchers. Communication channels should be secured with TLS certificates and robust authentication mechanisms; common protocols include IEC 61850, IEC 60870-5-104, and Modbus TCP.

\section{The Innosuisse Flagship STORE}
STORE is an R\&D project financed by Swiss Innovation Agency (Innosuisse) under the flagship program (Innosuisse No. 108.230 FS-EE, December 2023 to July 2027). The project addresses the design of a functional energy storage infrastructure for Switzerland to support 100\% production from renewable generation and energy self-sufficiency. The project is led by HES-SO Valais-Wallis and brings together ETHZ, EPFL, and BFH as academic partners, alongside a consortium of thirteen industrial partners spanning generation companies (Alpiq, KWO, Romande Energie), distribution and transmission system operators (OIKEN, RELL, Valgrid, Swissgrid), large industrial consumers and prsoumers (Novelis, Kieswerk AG Naters, FLASA), and technology specialists (Natron, Planair, Emissium).

The project is structured around three overarching research questions:
\begin{itemize}
    \item How should individual energy storage use cases be designed and operated to maximize value for their owners;
    \item How do these use cases collectively contribute to grid operations at the system level and which are the most relevant from an application perspective and financial sustainability perspective?
    \item  What energy storage infrastructure, in terms of technology mix, siting, sizing, and ramping capabilities, is required to achieve a reliable and cost-effective transition to 100\% renewables?
\end{itemize}

It is organized into the following-described five subprojects. 

\paragraph{Subproject 1: next-generation hydropower storage} develops sizing and control methods for hybrid energy storage systems combining hydropower plants with co-located BESS. Its objectives are to improve market participation, reduce mechanical wear from increased regulation duties, and mitigate hydropeaking, with KWO, Romande Energie, and Alpiq as industrial partners \cite{Drakaki2025,IdOmar2026}.

\paragraph{Subproject 2: integrated gas and electricity systems} investigates reversible solid oxide cell (rSOC) technology as a means of achieving seasonal energy storage through power-to-methane conversion, with demonstration use cases at OIKEN and Novelis targeting year-round renewable energy supply and industrial decarbonization respectively. A small-scale demonstrator has been built at the Energypolis campus in Sion as a joint effort between HES-SO Valais-Wallis and EPFL \cite{RTS2025PtG}.

\paragraph{Subproject 3: behind-the-meter energy storage} develops methods for jointly sizing and controlling storage at industrial and commercial consumer premises in a multi-service context, combining peak shaving, PV self-consumption, ancillary service provision, and, where applicable, hydrogen production for mobility markets. The use cases involve two industrial facilities, a gravel production site (Kieswerk Naters) and a textile factory (FLASA), both operating large behind-the-meter PV installations whose value they wish to maximize beyond current self-consumption practices.

\paragraph{Subproject 4: energy storage in distribution grids} addresses the use of storage as a non-wire alternative to grid reinforcement in low-voltage networks, developing both DSO-owned and DSO-coordinated behind-the-meter operational models. With OIKEN, results are being elaborated that, for the first time in the technical literature, address the hosting capacity of real distribution grids using data from more than 1,300 substations across low- and medium-voltage networks. With RELL, a control strategy is being developed for coordinating behind-the-meter assets to relieve network congestion, based on state-of-the-art optimal-power-flow methods \cite{Grammatikos2025}, with a direct link to the operator's control room.

\paragraph{Subproject 5: integration of energy storage in the bulk power system} takes a system-level perspective, formulating the coordinated siting and sizing of the national storage infrastructure as a large-scale optimization problem that minimizes total system cost subject to grid constraints, ancillary service requirements, and life cycle metrics. Its central planner function produces an optimal storage configuration that serves simultaneously as a technical benchmark for assessing market-driven deployment scenarios and as an input to the design of new electricity and ancillary service market rules \cite{Larroux2026,Jia2026}. A distinctive feature of this subproject is the availability of grid data from all operator levels, from transmission level 1 to low-voltage level 7, across the Central Valais region, which will enable a uniquely complete analysis of the question of at which aggregation level energy storage is best placed.
The project is currently in its core development phase and results are emerging progressively. They will be published on the STORE project website \url{http://flagship-store.ch} and in the specialized literature as they become available.

\section{Conclusion}
Energy storage is both a local asset and a system resource, and its full economic and technical value is only realized when individual deployments are designed and operated with awareness of the complete picture of the services it can provide. The taxonomy presented in this paper organizes energy storage applications across grid levels, ownership models, and operational objectives, providing a framework useful both for understanding existing deployments and for planning new ones.
For operators willing to develop an energy storage application, three interconnected implementation layers must be addressed: determining the right technology, size, and location at the design stage (and grid node, if relevant); managing the intertemporal dynamics of stored energy through scheduling; and governing the asset in real time within the constraints imposed by BESS with appropriate real-time control strategies integrated into the operator’s SCADA.
Finally, the STORE project was presented as a concrete example of how these challenges are addressed in practice across five application domains, with the ambition of providing operators, investors, and policymakers with the tools needed to integrate energy storage and their applications into the power grid.

\bibliographystyle{IEEEtran}
\small
\bibliography{references}

\end{document}